\documentstyle[preprint,aps]{revtex}
\begin{document}
\draft
\title{GENERALIZED RAYCHAUDHURI EQUATIONS FOR STRINGS AND MEMBRANES}
\author{Sayan Kar \thanks{ Electronic Address :
sayan@iopb.ernet.in} \thanks{ Address after Sept. 1, 1996 : IUCAA, Post
Bag 4, Ganeshkhind, Pune 411 007}
\thanks{Invited Talk at IAGRG XVIII, (Matscience, Madras, 15-17 Feb., 1996)}}
\address{Institute of Physics,\\
Sachivalaya Marg, Bhubaneswar--751005, INDIA}
\maketitle 
\begin{abstract}
A recent generalisation of the Raychaudhuri equations
for timelike geodesic congruences to families of $D$ dimensional
extremal, timelike, Nambu--Goto surfaces embedded in an $N$ dimensional 
Lorentzian background
is reviewed. Specialising to $D=2$ (i.e the case of
string worldsheets) we reduce the equation for the 
generalised expansion $\theta _{a}, (a =\sigma,\tau)$ to a second order,
linear, hyperbolic partial differential equation which resembles a 
variable--mass
wave equation in $1+1$ dimensions. Consequences, such as
a generalisation of geodesic focussing to families of worldsheets
as well as exactly solvable cases are explored and analysed in
some detail.
Several possible directions of future research are also pointed out.

\end{abstract}

\newpage

\section{INTRODUCTION}

It is a well established fact today that the proof of the
existence of spacetime singularities in the general theory
of relativity (GR) largely relies on the consequences
obtained from the Raychaudhuri equations for null/timelike
geodesic congruences [1,2,3]. Even though the applications of the
Raychaudhuri equations are mostly confined to the domain
of GR, it is important to note that these equations contain
some basic statements about the nature of geodesics in a
Riemannian/pseudo--Riemannian geometry. GR comes into the
picture when one assumes the Einstein field equation and
thereby reduces one crucial term containing information 
about geometry  into an object related to matter stress--energy.
Subsequently, if one imposes an Energy condition (such as the
Weak Energy condition which implies that the energy density of
matter is always positive in all frames of reference) it is
possible to derive the fact that geodesic congruences necessarily
converge within a finite value of the affine parameter. This
is known as the focussing theorem, which, along with other
assumptions about causality, essentially imply the
existence of spacetime singularities.
 
In string and membrane theories, the notion of the point particle
(and its associated world--line) which is basic to 
GR as well as other relativistic field theories,
gets replaced by the string/membrane (and its corresponding
world--surface). This is a radically new concept and has paid
rich dividends in recent times. For instance, it is claimed
that quantum gravity as well as a unification of forces
comes out naturally from quantizing string theories.

If one accepts the string/membrane viewpoint then it should
,in principle, be possible  to derive  
the corresponding generalized Raychaudhuri
equations for timelike/null worldsheet congruences and arrive at
similar focussing and singularity theorems in Classical String
theory.
Very recently, Capovilla and Guven [4] have written down the 
generalized Raychaudhuri equations for timelike worldsheet
congruences. In this talk, we  shall first give a brief review
of these equations. Thereafter, we construct explicit examples of these 
rather complicated set of equations by specializing to certain
simple extremal families of surfaces. Our principal
aim is to extract some information regarding {\em focussing of families
of surfaces} in a way similar to the results for geodesic congruences 
in GR. 

Finally, we shall summarize some recent work in progress
and point out several future directions of research.

\section{REVIEW}

\subsection{What are Raychaudhuri Equations ?}

An useful way of visualising the content of the Raychaudhuri equations
is to look at an analogy with fluid flow. The flow lines of a fluid
(which are the integral curves of the velocity field) form a congruence
of curves. If we focus our attention on the cross--sectional area 
enclosing a certain number of these curves we find that it is
different at different points. The gradient of the velocity field
contains crucial information about the behaviour of this area.
What can happen to this area as we move along the family of curves?
The obvious answer is--it can {\em expand} (i.e. a smaller circle may
become a larger circle which is concentric to the former), it can
become {\em sheared} (i.e. a circle can become an ellipse) or it can
twist or {\em rotate}. The information about each of these objects
(i.e. the expansion, shear and the rotation) is encoded in the 
gradient of the velocity field). Recall that this gradient is a
second rank tensor. Such a quantity can be split into its
symmetric traceless, antisymmetric and trace parts -- these are
respectively the shear, rotation and expansion of the congruence
/flow lines. For geodesic congruences in Riemannian/pseudo--Riemannian
geometry we have to replace the flow lines with geodesics and the
velocity field by the tangent vectors to the geodesic curves.
  
Therefore, expressed in mathematical language we have :

\begin{equation}
v_{\mu ; \nu} = \sigma_{\mu\nu} + \omega_{\mu\nu} + \frac{1}{3}\theta 
h_{\mu\nu}
\end{equation}

The {\em evolution} equation for each of these quantities -- expansion,
shear, rotation along the geodesic congruence
are what are known as the Raychaudhuri equations. 

For example, the equation for the expansion $\theta$ for a timelike geodesic
congruence turns out to be :

\begin{equation}
\frac{d\theta}{d\lambda} + \frac{1}{3}\theta^{2} + \sigma^{2} -\omega^{2}
= -R_{\mu\nu}\xi^{\mu}\xi^{\nu}
\end{equation}

An analysis of the nature of the solutions of this equation leads us to
the concept of geodesic focussing. First, let us convert this equation into
a second order linear equation by redefining $\theta = 3\frac{F^{\prime}}{F}$.
This results in the following :

\begin{equation}
\frac{d^{2}F}{d\lambda^{2}} + \frac{1}{3}
\left ( \sigma^{2} -\omega^{2} + R_{\mu\nu}
\xi^{\mu}\xi^{\nu} \right ) F = 0
\end{equation}

Assuming zero rotation, one can prove that if $\theta$ is negative somewhere
it has to go to $-\infty$ within a finite value of the affine parameter
if the coefficient of the second term in the above equation is
greater than or equal to zero. 
This largely follows from the theorems on the existence of zeros
of a general class of such second order, linear equations proofs of which
can be found in the paper by Tipler [6].

What happens when $\theta\rightarrow -\infty$? If $A_{1}$ is the area at
$\lambda_{1}$ and $A_{2}$ at $\lambda_{2}$ we can write $\theta$ as
$\theta = \frac{A_{2} - A_{1}}{A_{2}}$. Therefore $\theta$  can go to
$-\infty$ if $A_{2}\rightarrow 0$. Thus, the family of geodesics must
focus at a point (i.e. at $\lambda = \lambda_{2}$).

A word about the constraints on the geometry which are necessary for
focussing. Notice that if the rotation is zero we must have 
$R_{\mu\nu}\xi^{\mu}\xi^{\nu} \ge 0$. Using the Einstein field
equations one can write this as $T_{\mu\nu} - \frac{1}{2}g_{\mu\nu}T \ge 0$.
This is what is known as the Strong Energy Condition. An Energy Condition
, in general, defines a certain class of matter which is physically
possible. For instance, the Weak Energy Condition states that
$T_{\mu\nu}\xi^{\mu} \xi^{\nu}\ge 0$ which physically implies
the positivity of energy density in all frames of reference.

We now ask the following questions:

Suppose we replace extremal curves (geodesics) by extremal
$D$ dimensional timelike surfaces embedded in an $N$ dimensional
background.
\begin{center}
\begin{itemize}
\item{Are there  generalisations of the Raychaudhuri equations?}
\item{Is there an analog of geodesic focussing?}
\end{itemize}
\end{center}

The remaining part of this article is devoted to answering these two
questions.
\subsection{Geometry of Embedded Surfaces}

We begin with a brief account of the differential geometry
of embedded surfaces.

A $D$ dimensional surface in an $N$ dimensional background is defined through
the embedding $x^{\mu} = X^{\mu}(\xi^{a})$ where $\xi^{a}$ are the coordinates
on the surface and $x^{\mu}$ are the ones in the background. Furthermore, 
we construct an orthonormal basis  $(E^{\mu}_{a}, n^{\mu}_{i})$ consisting 
of $D$ tangents and $N-D$
normals at each point on the surface. $E^{\mu}_{a}$, $n^{\mu}_{i}$ satisfy
the following properties.

\begin{equation}
g(E_{a}, E_{b}) = {\eta}_{ab} \quad;\quad g(E_{a},n_{i}) = 0 \quad;
\quad g(n_{i},n_{j}) = {\delta}_{ij}
\end{equation} 

We can write down the Gauss--Weingarten equations using the 
usual definitions of extrinsic curvature, twist potential and the 
worldsheet Ricci rotation coefficients.

\begin{eqnarray}
D_{a}E_{b} = {\gamma}_{ab}^{c}E_{c} -K _{ab}^{i}n_{i} \\
D_{a}n_{i} = K_{ab}^{i}E^{b} + {\omega}^{ij}_{a}n_{j}
\end{eqnarray}

where $D_{a} \equiv E^{\mu}_{a} D_{\mu}$ ( $D_{\mu}$ being the usual 
spacetime covariant derivative). The quantities $K_{ab}^{i}$ (extrinsic 
curvature), ${\omega}_{a}^{ij}$ and
 ${\gamma}_{ab}^{c}$ are defined as :

\begin{eqnarray}
K_{ab}^{i} = -g(D_{a}E_{b}, n^{i}) = K_{ba}^{i} \\
{\omega}_{a}^{ij} = g(D_{a}n^{i},n^{j})\\
{\gamma}_{abc} = g(D_{a}E_{b}, E_{c}) = -{\gamma}_{acb}
\end{eqnarray}

In order to analyse deformations normal to the worldsheet we need to
consider the normal gradients of the spacetime basis set.
The corresponding analogs of the Gauss--Weingarten
equations are :

\begin{eqnarray}
D_{i}E_{a} = J_{aij}n^{j} + S_{abi}E^{b} \\
D_{i}n_{j} = -J_{aij}E^{a} + {\gamma}^{k}_{ij}n_{k}
\end{eqnarray}

where $D_{i} \equiv n^{\mu}_{i} D_{\mu}$ .
The quantities $J_{a}^{ij}$, $S_{abi}$ and
 ${\gamma}^{k}_{ij}$ are defined as :

\begin{eqnarray}
S^{i}_{ab} = g(D^{i}E_{a}, E_{b}) = -S^{i}_{ba}\\ 
{\gamma}_{ijk} = g(D_{i}n_{j}, n_{k}) = -{\gamma}_{ikj}\\
J_{a}^{ij} = g(D^{i}E_{a},n^{j}) 
\end{eqnarray}

\subsection{Sketch of Derivation}

The full set of equations governing the evolution of deformations
can now be obtained by taking the worldsheet gradient of $J_{a}^{ij}$.
This turns out to be (for details see Appendix of Ref. [4]).

\begin{equation}
{\tilde{\nabla}}_{b}J^{ij}_{a} = -{\tilde{\nabla}}^{i}K_{ab}^{j}
-{J_{b}^{i}}_{k}{J_{a}^{kj}} - K_{bc}^{i}K^{cj}_{a} - g(R(E_{b},n^{i})E_{a},
n^{j})
\end{equation}

where the extrinsic curvature tensor components are $K_{ab}^{i} =
-g_{\mu\nu}E^{\alpha}_{a}(D_{\alpha}E^{\mu}_{b})n^{\nu i}$

On tracing over worldsheet indices we get
\begin{equation}
{\tilde{\nabla}}_{a}J^{aij} = -{J_{a}^{i}}_{k}{J^{akj}} - K_{ac}^{i}K^{acj}
 - g(R(E_{a},n^{i})E^{a},
n^{j})
\end{equation}

where we have used the equation for extremal membranes (i.e. $K^{i} = 0$)

The antisymmetric part of (12) is given as :

\begin{equation}
{\tilde{\nabla}}_{b}J^{ij}_{a}- {\tilde{\nabla}}_{a}J^{ij}_{b}
= G_{ab}^{ij}
\end{equation}

where $g(R(Y_{1},Y_{2})Y_{3},Y_{4}) = R_{\alpha\beta\mu\nu}Y_{1}^{\alpha}
Y_{2}^{\beta}Y_{3}^{\mu}Y_{4}^{\nu}$ and

\begin{equation}
G_{ab}^{ij} = -{J_{b}^{i}}_{k}{J_{a}^{kj}} - K_{bc}^{i}K^{cj}_{a} - g(R(E_{b},n^{i})E_{a},
n^{j}) - (a\rightarrow b)
\end{equation}

One can further split $J_{aij}$ into its symmetric traceless, trace and
antisymmetric parts ($J_{a}^{ij} = {\Sigma}_{a}^{ij} + {\Lambda}_{a}^{ij}
+ \frac{1}{N-D}{\delta}^{ij}{\theta_{a}}$) and obtain the evolution equations
for each of these quantities. The one we shall be concerned with mostly
is given as 

\begin{equation}
\Delta \gamma + {\frac{1}{2}}{\partial}_{a}\gamma{\partial}^{a}\gamma + 
(M^{2})^{i}_{i} = 0
\end{equation} 

with the quantity $(M^{2})^{ij}$ given as :

\begin{equation}
(M^{2})^{ij} = K_{ab}^{i}K^{abj} +
R_{\mu\nu\rho\sigma}E^{\mu}_{a}n^{\nu i}E^{\rho a}n^{\sigma j}
\end{equation} 

 ${\nabla}_{a}$ is the worldsheet covariant derivative ($\Delta = {\nabla^{a}
 \nabla_{a}}$) and ${\partial}_{a}\gamma={\theta}_{a}$. 
  Notice that 
 we have set ${\Sigma}_{a}^{ij}$ and 
${\Lambda}^{ij}_{a}$ equal to zero. This is possible only if the symmetric 
traceless part of $(M^{2})^{ij}$ is zero. One can check this by looking 
at the full set of generalized Raychaudhuri equations involving 
$\Sigma^{ij}_{a}$, ${\Lambda}^{ij}_{a}$ and $\theta_{a}$ [4].
For geodesic curves the usual Raychaudhuri
equations can be obtained by noting that $K^{i}_{00} = 0$, the $J_{aij}$ are 
related to their spacetime counterparts $J_{\mu\nu a}$ through
the relation $J_{\mu\nu a} = n^{i}_{\mu}n^{j}_{\nu}J_{aij}$,
and the $\theta$ is defined by contracting with the projection 
tensor $h_{\mu\nu}$.

The $\theta _{a}$ or $\gamma$ basically tell us how the
spacetime basis vectors change along the normal directions  as we
move along the surface. If ${\theta}_{a}$ diverges somewhere
, it induces a divergence in $J_{aij}$ , which, in turn means
that the gradients of the spacetime basis along the normals have
a discontinuity. Thus the family of worldsheets meet along a
curve and a cusp/kink is formed. This, we claim, is a  
 focussing effect for extremal surfaces analogous to 
 geodesic focussing in GR where families of geodesics focus at a point 
 if certain specific conditions on the matter stress energy are
 obeyed.

\subsection{Is the evolution of $J_{aij}$ constrained ?}

For each pair ($ij$) there are $D$ quantities $J_{a}^{ij}$.
To analyse whether the evolution of $J^{ij}_{a}$ is constraint--free
or not we split the antisymmetric equations into two sets.

\begin{eqnarray}
{\tilde{\nabla}}_{0}J^{ij}_{A}- {\tilde{\nabla}}_{A}J^{ij}_{0}
= G_{ab}^{ij}\\
{\tilde{\nabla}}_{B}J^{ij}_{A}- {\tilde{\nabla}}_{A}J^{ij}_{B}
= G_{ab}^{ij}
\end{eqnarray}

Thus first of these contains a total of $D$ equations
for each $ij$. Eqn (12) on the other hand contains one equation.
Thus the total number of equations contained in these is
$D$ which is the number necessary to specify the evolution of
the quantity $J_{a}^{ij}$. Therefore the second set above is
actually a set of constraints on the evolution.

However, note that the number of equations in this second set
is equal to $\frac{1}{2}(D-1)(D-2)$. Therefore for $D=1$ (curves)
and $D=2$ (string worldsheets) these equations are vacuous and the
evolution of $J_{a}^{ij}$ is entirely constraint--free!
Even more surprising is the fact that for any general $D>2$ also
the second set of equations are {\em identities} (for a proof see
[4]). Therefore, if the constraints and equations of motion are 
satisfied at any initial time they continue to be so for all 
future values.

\section{FOCUSSING OF STRING WORLDSHEETS [5]}
 
Two dimensional timelike surfaces embedded in a four dimensional
background are the objects of discussion in this section. We begin
by writing down the generalised Raychaudhuri equation  for the case
in which $\Sigma_{a}^{ij}$ and $\Lambda_{a}^{ij}$ are set to zero
(i.e implicitly assuming that $(M^{2})^{ij}$ does not have a nonzero
symmetric traceless part.Thus we have 

\begin{equation}
- \frac{{\partial}^{2}F}{\partial{\tau}^{2}} +
\frac{{\partial}^{2}F}{\partial{\sigma}^{2}}
+\Omega^{2}(\sigma,\tau)  (M^{2})^{i}_{i}(\sigma,\tau)F = 0
\end{equation}

where ${\Omega}^{2}$ is the conformal factor of the induced 
metric written in isothermal coordinates.
Notice that the above equation is a second--order, linear, hyperbolic
partial differential equation. On the contrary, the Raychaudhuri equation for
curves is an linear, second order, ordinary differential equation. 
The easiest way to analyse the solutions of this 
equation is to assume separability of the 
quantity $\Omega^{2}(M^{2})^{i}_{i}$. Then, we have 

\begin{eqnarray}
\Omega^{2}(M^{2})^{i}_{i} = M^{2}_{1}(\tau) + M^{2}_{2}(\sigma)\\
F(\tau,\sigma) = F_{1}(\tau) \times F_{2}(\sigma)
\end{eqnarray}

With these we can now split the partial differential equation
into two ordinary differential equations given by 

\begin{eqnarray}
\frac{d^{2}F_{1}}{d\tau^{2}} + (\omega^{2} - M^{2}_{1}(\tau))F_{1} = 0 \\
\frac{d^{2}F_{2}}{d\sigma^{2}} + (\omega^{2} + M^{2}_{2}(\sigma))F_{2} = 0
\end{eqnarray}

Since the expansions along the $\tau$ and $\sigma$ directions can be
written as $\theta_{\tau} = \frac{\dot F_{1}}{F_{1}}$ and $\theta_{\sigma}
=\frac{F_{2}^{\prime}}{F_{2}}$ we can analyse focussing effects by 
locating the zeros of $F_{1}$ and $F_{2}$ much in the same way as for
geodesic curves [6]. 
The well--known theorems on the existence of zeros 
of ordinary differential equations as discussed in [6] 
make our job much simpler. The theorems essentially state that the solutions
of equations of the type $\frac{d^{2}A}{dx^{2}} + H(x)A = 0$ will have 
at least one zero iff $H(x)$ is positive definite.
Thus for our case here, we can conclude that, focussing 
along the $\tau$ and $\sigma$ directions will take place only if

\begin{equation}
\omega^{2} \ge max[M^{2}_{1}(\tau)]
\qquad ; \qquad \omega^{2} \ge max[-M^{2}_{2}(\sigma)]
\end{equation}

 For stationary strings, one notes that $(M^{2})^{ij}$ will
not have any dependence on $\tau$. Thus we can set $M_{1}^{2}$ equal to
zero. Thus, focussing will entirely depend on the sign of the quantity
$M^{2}_{2}$. We can write $M^{2}_{2}$ alternatively as follows. Consider the
Gauss--Codazzi integrability condition:

\begin{equation}
R_{\mu\nu\alpha\beta}E^{\mu}_{a}E^{\nu}_{b}E^{\alpha}_{c}E^{\beta}_{d}
= R_{abcd} - K_{aci}K^{aci} + K_{bdi}K^{bdi}
\end{equation}

Trace the above expression on both sides with $\eta^{ac}\eta^{bd}$ and 
reaarange terms to obtain :

\begin{equation}
K_{abi}K^{abi} = -^{2}R + K^{i}K_{i} 
+ R_{\mu\nu\alpha\beta}E^{\mu}_{a}E^{\alpha a}
E^{\nu}_{b}E^{\beta b}
\end{equation}

Thereafter, use this expression and the fact that $n^{i\mu}n^{\nu}_{i}
= g^{\mu\nu} - E^{\mu a}E^{\nu}_{a}$ in the original expression for
$(M^{2})^{i}_{i}$ (see Eqn.(11)) to get

\begin{equation}
M^{2}_{2} = - {}^{2}R + R_{\mu\nu}E^{\mu a}E^{\nu}_{a}
\end{equation}

One can notice the following features from the above expression:

(i) If the background spacetime is a vacuum solution of the Einstein 
equations then the positivity of  $M^{2}_{2}$ is guaranteed iff
$^{2}R \le  0$. Thus all string configurations in vacuum spacetimes
which have negative Ricci curvature everywhere will necessarily imply
focussing. This includes the well known string solutions in
Schwarzschild and Kerr backgrounds.

(ii) If the background spacetime is a solution of the Einstein equations
then we can replace the second and third terms in the expressions
for $M^{2}_{2}$ by the corresponding ones involving the Energy
Momentum tensor $T_{\mu\nu}$ and its trace. Thus we have 

\begin{equation}
M^{2}_{2} =\left( - \frac{1}{2}g_{\mu\nu}{}^{2}R + T_{\mu\nu} - \frac{1}{2}
Tg_{\mu\nu}\right )E^{\mu a}E^{\nu}_{a} 
\end{equation}

Notice that if we split the quantity $E_{a}^{\mu}E^{\nu a}$ into two terms
such as $E_{\tau}^{\mu}E^{\nu \tau}$ and $E_{\sigma}^{\mu}E^{\nu \sigma}$
then we have :

\begin{equation}
M^{2}_{2} = - ^{2}R + \left (T_{\mu\nu} - \frac{1}{2}
Tg_{\mu\nu}\right )E^{\mu \tau}E^{\nu}_{\tau} + \left (T_{\mu\nu} - \frac{1}{2}
Tg_{\mu\nu}\right )E^{\mu \sigma}E^{\nu}_{\sigma}
\end{equation}
 The second term in the above equation is the L. H. S. of the 
 Strong Energy Condition (SEC).
Apart from this we have two other terms which are entirely dependent on the 
fact that we are dealing with extended objects. The positivity of the
whole quantity can therefore be thought of as an {\em Energy Condition}
for the case of strings. Thus even if the background spacetime
satisfies the SEC, focussing of string world--sheets is not
guaranteed--worldsheet curvature and the projection of the combination
$T_{\mu\nu} - \frac{1}{2}g_{\mu\nu}T$ along the $\sigma$
direction have an important role to play in deciding  focussing/defocussing. 

Let us now try to understand the consequences of the above equations for 
certain specific flat and curved backgrounds for which the string solutions
are known.

\subsection{Rindler Spacetime}

The metric for four dimensional Rindler spacetime is given as 

\begin{equation}
ds^{2} = -a^{2}x^{2}dt^{2} + dx^{2} + dy^{2} + dz^{2}
\end{equation}

We recall from [7] the a string solution in a Rindler spacetime:

\begin{eqnarray}
t=\tau \quad ; \quad x = ba\cosh {a\sigma_{c}} \quad ;{\nonumber} \\ \quad 
y = ba^{2}\sigma_{c} \quad ; \quad z = z_{0} \quad (constant) 
\end{eqnarray}

where $d\sigma_{c}=\frac{d\sigma}{{a^{2}x^{2}}}$ and $b$ is an integration
constant.
The orthonormal set of tangents and normals to the worldsheet  can be chosen
to be as  follows:

\begin{equation} 
E_{\tau}^{\mu} \equiv \left ( \frac{1}{ax}, 0, 0, 0 \right ) \quad ; \quad
E_{\sigma}^{\mu} \equiv \left ( 0, \tanh {a\sigma_{c}}, sech {a\sigma_{c}}, 0
\right )
\end{equation}

\begin{equation}
n^{\mu}_{1} \equiv \left ( 0, 0, 0, 1 \right ) \quad ; \quad 
n^{\nu}_{2} \equiv \left ( 0, sech a\sigma_{c}, - \tanh a\sigma_{c}, 0 \right )
\end{equation}

In the worldsheet coordinates $\tau , \sigma_{c}$ the induced metric is flat
and the components of the extrinsic curvature tensor turn out to be

\begin{eqnarray}
K^{1}_{ab} = 0 \quad ; \quad K^{2}_{\tau\tau} = -K^{2}_{\sigma_{c}\sigma_{c}}
 = \frac{1}{ba \cosh^{2}a\sigma_{c}} \quad ; {\nonumber} \\
   K^{2}_{\sigma\tau} = 0
\end{eqnarray}

The quantity $(M^{2})^{i}_{i}$ which is dependent only on the extrinsic curvature
of the worldsheet (the background spacetime being flat) turns out to be

\begin{equation}
(M^{2})_{i}^{i} = \frac{2}{b^{2}a^{2}\cosh^{4}a\sigma_{c}}
\end{equation}

Therefore the generalized Raychaudhuri equation turns out to be

\begin{equation}
-\frac{{\partial}^{2}F}{{\partial}\tau^{2}} +
\frac{{\partial}^{2}F}{{\partial}\sigma_{c}^{2}} + \frac{2a^{2}}{\cosh^{2}
a\sigma_{c}} F
= 0
\end{equation}

Separating variables ($F= T(\tau)\Sigma(\sigma)$ ) one gets the harmonic oscillator
equation for $T$ and the Poschl Teller equation for positive eigenvalues 
for $\Sigma$ which is given as:

\begin{equation}
\frac{d^{2}\Sigma}{d\sigma^{2}} + \left ( {\omega}^{2} +
\frac{2a^{2}}{\cosh^{2}\sigma} \right ) \Sigma = 0
\end{equation} 

From the results of Tipler [6] on the zeros of differential equations
one can conclude that focussing will occur ($H(\sigma) > 0$ always). 

Several other examples can be found in [5] and [9].

\section{FOCUSSING OF HYPERSURFACES [5]}

We now move on to the special case of timelike hypersurfaces. 
Here we have $D$ quantities $J_{a}$ but only one normal
defined at each point on the surface. The Eqn. (8) turns out to 
be :

\begin{equation}
{\partial}_{b}J_{a} - {\partial}_{a}J_{b} = 0
\end{equation}

Therefore one can write $J_{a} = {\partial}_{a}\gamma$ and the traced 
equation (7) becomes,

\begin{equation}
\Delta \gamma + ({\partial}_{a}\gamma )( {\partial}^{a}{\gamma}) + M^{2} = 0
\end{equation}

with

\begin{eqnarray} 
M^{2} =  K_{ab}K^{ab} + R_{\nu\sigma}n^{\nu}n^{\sigma} \nonumber \\
=-^{2}R + R_{\mu\nu}E^{\mu a}E^{\nu}_{a}
=-^{2}R + ^{3}R - R_{\mu\nu}n^{\mu}n^{\nu}
\end{eqnarray}

where we have used $n^{\mu}n^{\nu} = g^{\mu\nu} - E^{\mu}_{a}E^{\nu a}$
and the Gauss--Codazzi integrability condition.

If we assume that the background spacetime satisfies the Einstein equation
then we have:

\begin{equation}
M^{2} = -\left ( ^{2}Rg_{\mu\nu} + T_{\mu\nu} + Tg_{\mu\nu}\right )
n^{\mu}n^{\nu}
\end{equation}

Thus, for stationary two dimensional hypersurfaces (strings in 3D backgrounds)
we have the same conclusions as obtained in the previous section.  
For a two--dimensional hypersurface in  three--dimensional 
flat background the task is even simpler. $M^{2}$ can be shown to be equal to
the negative of the Ricci scalar of the membrane's induced metric and 
$^{2}R\leq 0$ guarantees focussing. 
 
Let us now turn to a specific case where the equations are exactly solvable.
 
\subsection{Hypersurfaces in a $2+1$ Curved Background}

 Our backgound spacetime here is curved, 
Lorentzian background and $2+1$ dimensional. The metric we choose is that
of a Lorentzian wormhole in $2+1$ dimensions given as :

\begin{equation}
ds^{2} = -dt^{2} + dl^{2} + \left ( b_{0}^{2} + l^{2}\right ) d\theta^{2}
\end{equation}

A string configuration in this background can be easily found by solving the 
geodesic equations in the $2D$ spacelike hypersurface [8]. This turns out to be

\begin{equation}
t=\tau \quad ; \quad l=\sigma \quad ; \quad \theta = \theta_{0}
\end{equation}

The tangents and normal vectors are simple enough:

\begin{equation}
E^{\mu}_{\tau} \equiv \left ( 1, 0, 0  \right )  \quad ; \quad 
E^{\mu}_{\sigma} \equiv \left ( 0, 1, 0 \right ) \quad ; \quad 
n^{\mu} = \left ( 0, 0, \frac{1}{b^{2}_{0} + l^{2}} \right )
\end{equation}

The extrinsic curvature tensor components are all zero as the induced metric
is flat. Using the Riemann tensor components (which can be evaluated simply
using the standard formula) we can write down the generalised Raychaudhuri
equation. This turns out to be :

\begin{equation}
-\frac{{\partial}^{2}F}{{\partial}\tau^{2}} +
\frac{{\partial}^{2}F}{{\partial}\sigma^{2}} 
+ \left (-\frac{b_{0}^{2}}{(b_{0}^{2} + \sigma^{2})^{2}} \right ) F = 0
\end{equation}  

A separation of variables $F= T(\tau)\Sigma(\sigma)$ will result in two
equations--one of which is the usual Harmonic Oscillator and the other 
given by:

\begin{equation}
\frac{d^{2}\Sigma}{d\sigma^{2}} + \left ( \omega^{2} - \frac{b_{0}^{2}}{(b_{0}
^{2} + \sigma^{2})^{2}} \right ) \Sigma = 0 
\end{equation}

The above equation can be recast into the one for Radial Oblate Spheroidal Functions 
by a simple change of variables -- $ \Sigma^{\prime} = \sqrt{b_{0}^{2} + 
\sigma^{2}} \Sigma $.

\begin{equation}
(1+ \xi^{2} ) \frac{d^{2}{\Sigma^{\prime}}}{d\xi^{2}} +
 2\xi\frac{d\Sigma^{\prime}}{d\xi} + \left (\omega^{2}b_{0}^{2}(1+\xi^{2}) \right )
 \Sigma^{\prime} = 0
 \end{equation}

where $\xi =\frac{\sigma}{b_{0}}$.

 The general equation for Radial Oblate Spheroidal Functions is given as :
 
 \begin{equation}
(1+ \xi^{2} ) \frac{d^{2}{V_{mn}}}{d\xi^{2}} +
 2\xi\frac{dV_{mn}}{d\xi} + \left (-\lambda_{mn} + 
 k^{2}{\xi}^{2}  
 - \frac{m^{2}}{1+\xi^{2}} \right ) V_{mn} = 0
 \end{equation}

Assuming $m=0$ and $ \lambda_{0n} = -k^{2} = -\omega^{2}b_{0}^{2} $
we get the equation for our case. The solutions are finite at infinity 
and behave like simple sine/cosine waves in the variable $\sigma$. 
Consulting the tables in [10] we conclude that only for $n=0,1$ we 
can have $\lambda_{0n}$ to be negative. In general, the scattering 
problem for the Schroedinger--like equation has been analysed 
numerically in [11].
    
As regards focussing, one can say from the differential equations and the 
theorems stated in [5] that the function $\Sigma^{\prime}$ will always have 
zeros if $\omega^{2} \ge \frac{1}{b_{0}^{2}}$. Even from the series 
representations (see [9]) of the Radial Oblate Spheroidal Functions we 
can exactly locate the zeros and obtain explicitly the focal curves. However,
we shall not attempt such a task here.

\section{WORK IN PROGRESS AND FUTURE DIRECTIONS}

A systematic study of the generalised Raychaudhuri equations has only
begun. A large number of open problems therefore exist in this 
subfield. Here we report briefly on some recently obtained results
and list a few of the outstanding issues.

(i) Although we have been able to derive an analog of geodesic focussing
for extremal timelike membranes by looking at some special cases
a general treatment of the problem is still lacking. For example,
recall that we made the crucial assumption of separability in the
equation for the case of strings and hypersurfaces. In order to
avoid this assumption, a way out could be to define an initial value 
problem with the expansions in different directions taking on a
negative value at specific locations on the surface. The initial
value as well as the partial differential equation can be recast
into one Volterra integral equation of the second kind. Solving this 
would then give the necessary condition for focussing. Note that in this
approach one does not assume an Energy Condition at the outset because
one does not have a choice to do so. Some progress along these directions
are to be reported in {\cite{sk:961}}.

(ii) What does one need to have an energy condition in this case?
Recall that in the Raychaudhuri equation for geodesic congruences
the appearance of the term $R_{\mu\nu}{\xi}^{\mu}{\xi}^{\nu}$ and the
Einstein equation relating geometry to matter were the deciding factors.
One could translate the {\em purely geometric} term into a term
containing {properties of matter stress energy}. Focussing was a
consequence of assuming certain physically relevant features of matter.
To have such a situation in the case of strings we need to have an
{\em Einstein equation} in string theory. This may sound outrageous
but we will see why it need not be so.

General relativity as a theory has a unique feature in comparison
to all other theories that we know of. {\em The motion of test 
particles can be derived from the Einstein field equations}. We must
remember that this is a fact which is not true in all other theories.
For example, in electrodynamics the Lorentz force law {\em cannot} in any
way be derived from the Maxwell equations. 
The basic question therefore reduces to the following :

{\em What is the field equation from which one can arrive at the
equation of motion for test strings ?}

An answer to this question will help us in analysing worldsheet
focussing by assuming an Energy condition much in the same way
as one does it for geodesic congruences.

(iii) It is important to note that we have restricted ourselves
exclusively to extremal Nambu--Goto type membranes while
deriving the generalised Raychaudhuri equations and exploring
its consequences. What are the corresponding equations for
actions other than Nambu--Goto (rigidity corrections, presence of 
antisymmetric tensor fields, supersymmetric generalisations etc.)?
It has been found {\cite{sk:962}} that in the presence of 
antisymmetric tensor fields one has several nontrivialities
appearing. Firstly, one cannot set the $\Sigma_{aij}, \Lambda{aij}$
equal to zero and work with only the equation of the expansions.
If one sets $\Sigma_{aij}$ equal to zero then one has to identify the
components of $\Lambda_{aij}$ with the projections of $H_{\mu\nu\rho}$.
Therefore, we can now attribute a physical meaning to $\Lambda_{aij}$
by associating it with the background antisymmetric tensor field's
projections. Further work is clearly needed to understand the
consequences of actions other than Nambu--Goto and is in progress. 
  
(iv) Finally, we indicate a possible application in a totally
different area--the theory of biological (amphiphilic) membranes.
Deformations of these membranes ($D=2$ hypersurfaces in a $D=3$
Euclidean background) can be analysed using the same formalism
as presented here and may turn out to be useful in understanding
the fluctuations of these objects. The simplest case of the 
catenoidal membrane is discussed in detail in the appendix to
[5].
\vspace{.3in}

\section*{ACKNOWLEDGEMENT}
I thank B. R. Iyer and G. Date for inviting me to this meeting
and for the excellent hospitality provided at the Institute
of Mathematical Sciences, Madras.

\end{document}